\documentclass[creativecommons]{eptcs}
\providecommand{\event}{FOCLASA 2012} 
\usepackage{breakurl}             
\usepackage[english]{babel} 
\usepackage[utf8]{inputenc}
\usepackage[T1]{fontenc}
\usepackage{courier,graphicx,xspace,cite}
\usepackage{listings}
\usepackage{hyperref}
\usepackage{amssymb}

\title{A Type-Safe Model of Adaptive Object Groups \thanks{This research
      was done in the context of the EU project FP7-231620 \emph{HATS}: Highly
      Adaptable and Trustworthy Software using Formal Models
      (\url{http://www.hats-project.eu}).}}
\author{Joakim Bj{\o}rk  \institute{University of
      Oslo, Norway} \email{joakimbj@ifi.uio.no} 
\and
Dave Clarke
\institute{Katholieke Universiteit Leuven, Belgium}
\email{dave.clarke@cs.kuleuven.be}
\and 
Einar Broch Johnsen
 \institute{University of
      Oslo, Norway} \email{einarj@ifi.uio.no} 
\and
Olaf Owe
 \institute{University of
      Oslo, Norway} \email{olaf@ifi.uio.no} }

\def\titlerunning{A Type-Safe Model of Adaptive Object Groups}
\def\authorrunning{J. Bj{\o}rk, D. Clarke, E.~B. Johnsen \& O. Owe}

\def\absdisplaysize{\fontsize{9}{10}}

\lstdefinelanguage{ABS}{keywords=
{new,newgroup,data,type,def,case,of,this,thiscomp,in,null,joins,leaves,subtypeOf, wait,class,interface,extends,implements,if,else,await,get,Fut,return,skip,while,module,acquire,except,as, suspend,component}, sensitive=true, comment=[l]{//},
  morecomment=[s]{/*}{*/},
  morestring=[b]"}

\lstdefinestyle{absstyle}{
language=ABS,columns=fullflexible,
 		   mathescape=true,%
 		   showstringspaces=false,%
keywordstyle=\absdisplaysize\bfseries\ttfamily,
commentstyle=\sl\sffamily,%
basicstyle=\absdisplaysize\ttfamily,
inputencoding=utf8,
extendedchars,xleftmargin=2em
}

\lstnewenvironment{abs}[1][]{%
 \lstset{style=absstyle,#1}%
   \csname lst@SetFirstLabel\endcsname}
  {\csname lst@SaveFirstLabel\endcsname}

\newcommand{\Absdisp}[1]{\mbox{\lstinline[language=ABS,columns=fullflexible,mathescape=true,
keywordstyle=\absdisplaysize\bfseries\ttfamily,
basicstyle=\absdisplaysize\ttfamily]!#1!}}

\newenvironment{absinline}[1][]{%
\lstset{language=ABS,
		 showstringspaces=false,
		 columns=fullflexible,mathescape=true,%
         keywordstyle=\bfseries\ttfamily,commentstyle=\sl\sffamily,%
         basicstyle=\absdisplaysize\ttfamily,inputencoding=utf8,
         extendedchars,xleftmargin=2em,#1}}
{}

\lstnewenvironment{absexample}{
~\\
\noindent 
\lstset{style=absstyle,
basicstyle=\ttfamily\footnotesize,
framerule=0pt,
frame=tblr,
xleftmargin=4pt,
}}
{}

\lstnewenvironment{absexamplen}[1][]{
\lstset{style=absstyle,
basicstyle=\absdisplaysize\ttfamily,
frame=tblr,
captionpos=b,
xleftmargin=4pt,#1
}}
{}

\newcommand{\TYPE}[1]{\mbox{\sffamily #1}}
\newcommand{\many}[1]{\overline{#1}} 
\newcommand{\key}[1]{\mbox{\ttfamily\bfseries #1}}
\newcommand{\option}[1]{[#1]}
\newcommand{\sep}{\:|\:}
\newcommand{\comp}[1]{\key{Group}\langle #1\rangle}
\newcommand{\ok}{\key{ok}}
\newcommand{\effect}[1]{\langle #1\rangle}
\newcommand{\dom}{\textsl{dom}}
\newcommand{\nt}[1]{\mathit{#1}}
\newcommand{\text}[1]{\textbf{#1}}
\newcommand{\eval}[2]{#2(#1)}
\newcommand{\evalT}[2]{#2^{T}(#1)}
\newcommand{\evalV}[2]{#2^{V}(#1)}

\def\myttsize{\fontsize{8}{9}}

\newcommand{\nspacetyperule}[3]{ 
  \begin{array}{c} 
    \textsc{\scriptsize ({#1})} \\ 
    #2 \\
    \hline \\[-9pt]
    #3 
  \end{array}} 

\newcommand{\ntyperule}[3]{ 
  \begin{array}{c} 
    \textsc{\scriptsize ({#1})} \\ 
    #2 \\ 
    \hline
    #3 
  \end{array} }

\newcommand{\nredrule}[2]{ 
  \begin{array}{c} 
    \textsc{\scriptsize ({#1})} \\ 
    #2  
  \end{array}} 

\begin{document}
\maketitle

\begin{abstract}
  Services are autonomous, self-describing, technology-neutral
  software units that can be described, published, discovered, and
  composed into software applications at runtime. Designing software
  services and composing services in order to form applications or
  composite services requires abstractions beyond those found in
  typical object-oriented programming languages.  This paper explores
  service-oriented abstractions such as service adaptation, discovery,
  and querying in an object-oriented setting.  We develop a formal
  model of adaptive object-oriented groups which offer services to
  their environment. These groups fit directly into the
  object-oriented paradigm in the sense that they can be dynamically
  created, they have an identity, and they can receive method
  calls. In contrast to objects, groups are not used for structuring
  code.  A group exports its services through interfaces and
  relies on objects to implement these services. Objects may join
  or leave different groups. Groups may dynamically export new
  interfaces, they support service discovery, and they can be queried
  at runtime for the interfaces they support. We define an operational
  semantics and a static type system for this model of adaptive object
  groups, and show that well-typed programs do not cause
  method-not-understood errors at runtime.
\end{abstract}

\section{Introduction}

Good software design often advocates a loose coupling between the
classes and objects making up a system. Various mechanisms have been
proposed to achieve this, including programming to interfaces, object
groups, and service-oriented abstractions such as service discovery.
By programming to interfaces, client code can be written independently
of the specific classes that implement a service, using interfaces
describing the services as types in the program.  Object groups
loosely organize a collection of objects that are capable of
addressing a range of requests, reflecting the structure of real-world
groups and social organizations in which membership is dynamic
\cite{lea93groups}; e.g., subscription groups, work groups, service
groups, access groups, location groups, etc. Service discovery allows
suitable entities (such as objects) that provide a desired service to
be found dynamically, generally based on a query on some kind of
interface. An advantage of designing software using these mechanisms
is that the software is more readily adaptable. In particular, the
structure of the groups can change and new services can be provided to
replace old ones. The queries to discover objects are based on
interface rather than class, so the software implementing the
interface can be dynamically replaced by newer, better versions,
offering improved services.

This paper explores service-oriented abstractions such as service
adaptation, discovery, and querying in an object-oriented setting.
Designing software services and composing services in order to form
applications or composite services require abstractions beyond those
found in typical object-oriented programming languages.  To this end,
we develop a formal model of adaptive object-oriented groups that also
play the role of service providers for their environment.  These
groups can be dynamically created, they have identity, and they can
respond to methods calls, analogously with objects in the
object-oriented paradigm.  In contrast to objects, groups are not used
for executing code.  A group exports its services through interfaces
and relies on objects to implement these services.  From the
perspective of client code, groups may be used as if they were objects
by programming to interfaces. However, groups support service-oriented
abstractions not supported by objects. In particular, groups may
dynamically export new interfaces, they support service discovery, and
they can be queried at runtime for the interfaces they support. Groups
are loosely assembled from objects: objects may dynamically join or
leave different groups.  In this paper we develop an operational
semantics and a static type system for this adaptive group model based
on interfaces, interface queries, groups, and service discovery. The
type system ensures that well-typed programs do not cause
method-not-understood errors at runtime.

The paper is organized as follows. Section~\ref{sec:lang} presents the
language syntax and a small example. A type and effect system for the
language is proposed in Section~\ref{sec:typing} and an operational
semantics in Section~\ref{sec:opsem}.  Section~\ref{sec:typesafe}
defines a runtime type system and shows that the execution of
well-typed programs is type-safe. Section~\ref{sec:related} discusses
related work and Section~\ref{sec:conc} concludes the paper.


\section{A Kernel Language for Adaptive Object Groups}
\label{sec:lang}
We study an integration of service-oriented abstractions in an
object-oriented setting by defining a kernel object-oriented language
with a Java-like syntax, in the style of Featherweight Java
\cite{igarashi01}. In contrast to Featherweight Java, types are
different from classes in this language: interfaces describe services
as sets of method signatures and classes generate objects which
implement interfaces.  By programming to interfaces, the client need
not know how a service is implemented.  For this reason, the language
has a notion of group which dynamically connects interfaces to
implementations.  Groups are first-class citizens; they have
identities and may be passed around. An object may dynamically join a
group and thereby add new services to this group, extending the
group's supported interfaces.  Objects may be part of several groups.
Both objects and groups may join and leave groups, thereby migrating
their services between groups.  The kernel language considers
concurrent objects which interact by synchronous method calls.
Concurrent activities are triggered by instantiating classes with
\Absdisp{run} methods (similar to overriding the run method of Java's
Thread class).  This simple concurrency model is relevant for
service-oriented systems.

\subsection{The Syntax}
The syntax of the kernel language is given in Figure~\ref{fig:syntax}.
A type $T$ in the kernel language is either a basic type, an interface
describing a service, or a group of interfaces. A \emph{program} $P$
consists of a list $\many{IF}$ of interface declarations, a list
$\many{CL}$ of class declarations, and a main block $\{\many{T}\
\many{x}; s \}$. The main block introduces a scope with local
variables $\many{x}$ typed by the types $\many{T}$, and a sequence $s$
of program statements. We conventionally denote by $\many{x}$ a list
or set of the syntactic construct $x$ (in this case, a program
variable), and furthermore we write $\many{T}\ \many{x}$ for the list
of typed variable declarations $T_1\ x_1; \ldots; T_n\ x_n$ where we
assume that the length of the two lists $\many{T}$ and $\many{x}$ is
the same. The types $T$ are the basic type $\TYPE{Bool}$ of Boolean
expressions, the empty interface $\TYPE{Any}$, the names $I$ of the
declared interfaces, and group types $\comp{\many{I}}$ which state
that a group supports the set $\many{I}$ of interfaces. The use of
types is further detailed in Section~\ref{sec:typing}, including the
subtyping relation and the type system.

\emph{Interface declarations} $IF$ associate a name $I$ with a set of
method signatures. These method signatures may be inherited from other
interfaces $\many{I}$ or they may be declared directly as $\many{Sg}$.
A method \emph{signature} $Sg$ associates a return type $T$ with a
name $m$ and method parameters $\many{x}$ with declared types
$\many{T}$.

\emph{Class declarations} $CL$ have the form $\Absdisp{class}\
C(\many{T}\ \many{x})\, \Absdisp{implements}\ \many{I}\
\{\,\many{T_1}\ \many{x_1};\{\many{T_2}\ \many{x_2}; s\};\,
\many{M}\}$ and associates a class name $C$ to the services declared
in the interfaces $\many{I}$. In $C$, these services are realized
using methods to manipulate the fields $\many{x_1}$ of types
$\many{T_1}$.  The constructor block $\{\many{T_2}\ \many{x_2}; s\}$
initializes the fields, based on the actual values of the formal class
parameters $\many{x}$ of types $\many{T}$. Remark that the
  constructor block is executed \emph{asynchronously}. Consequently,
  it can be used to trigger concurrent activities starting in a new
  instance of a class. The methods $M$ have a signature $Sg$ and a
method body $ \{ \many{T}\ \many{x};\ s;\Absdisp{return}\ x;\, \}$
which introduces a \emph{scope} with local variables $\many{x}$ of
types $\many{T}$ where the sequence of statements $s$ is executed,
after which the expression $e$ is returned to the client.

The \emph{expressions} $e$ of the kernel language consist of Java-like
expressions for reading program variables $x$, method calls
$x.m(\many{x})$ where the actual method parameters are given by
$\many{x}$, and object creation $\Absdisp{new} \ C(\many{x})$ where
the actual constructor parameters are given by $\many{x}$. Method
calls are synchronous and in contrast to Java all method calls are
synchronized; i.e., a caller blocks until a method returns and a
callee will only accept a remote call when it is idle. For simplicity,
the kernel language supports self-calls but not re-entrance (which
could be addressed using thread identities as in Featherweight Java
\cite{igarashi01}).  In addition, we consider two expressions which
are related to service-oriented software: $\Absdisp{newgroup}$
dynamically creates a new, empty group which does not offer any
services to the environment.  \emph{Service discovery} may be
localized to a named group $y$: the expression $\Absdisp{acquire}\ I\
\Absdisp{in}\ y\ \Absdisp{except}\ \many{x}$ finds some group $g$ or
object $o$ such that $g$ or $o$ offers a service better than $I$ (in
the sense of subtyping) and such that $g$ or $o$ is not in the set
$\many{x}$. If the $\Absdisp{in}\ y$ clause is omitted, then the
service provider $g$ or $o$ may be found anywhere in the system.

The \emph{statements} $s$ of the kernel language include standard
statements such as $\Absdisp{skip}$, assignments $x=e$, sequential
composition $s_1;s_2$, conditionals, and $\Absdisp{while}$-loops. To
simplify the kernel language, we keep a flat representation of
expressions; i.e., expressions must be assigned to program variables
before they can be used in other statements.  Service interfaces
$\many{I}$ are \emph{dynamically exported} through a group $y$ by the
expression $x~\Absdisp{joins}\ y~\Absdisp{as}\ \many{I}$, which states
that object or group $x$ is used to implement the interfaces
$\many{I}$ in the group $y$. Consequently, $y$ will support the
interfaces $\many{I}$ after $x$ has joined the group.  Objects and
groups $x$ may try to withdraw service interfaces $\many{I}$ from a
group $y$ by the expression $x\ \Absdisp{leaves}\ y ~\Absdisp{as}\
\many{I}\ \{ s_1 \}\ \Absdisp{else}\ \{ s_2 \}$. Withdrawing
interfaces from a group can lead to runtime exceptions which need to
be handled either by the client or by the service provider. In our
approach, the exception is handled on the server side; i.e.,
withdrawing interfaces $\many{I}$ from $y$ only succeeds if $y$
continues to offer all the interfaces of $\many{I}$, exported by other
objects or groups.  Thus, removals may not affect the type of $y$.  If
the removal is successful then branch $s_1$ is taken, otherwise $s_2$
is taken.  In addition, the language includes the statement $x\
\Absdisp{subtypeOf}\ I\ y\ \{ s_1 \}\ \Absdisp{else}\ \{ s_2 \}$ which
is used to \emph{query} a known group $x$ about its supported
interfaces. The statement works like a conditional and branches the
execution depending on whether the query succeeds or not.  If $x$
offers an interface better than $I$, the expanded knowledge of the
group $x$ becomes available through the variable $y$ in the scope of
the statements $s_1$. If $x$ does not offer an interface as good as
$I$, the branch $s_2$ is taken. Remark the introduction of a new name
for the group inside the scope, which ensures that the knowledge of
the extended type is local.  (By syntactic sugar, the variable $y$
need not appear in the surface syntax).

\begin{figure}[t!]
\myttsize
$$\begin{array}{cl}
      \begin{array}[t]{c@{\,:\,}l}
         \multicolumn{2}{l}{\emph{Syntactic Categories.}}\\
        C&\TYPE{Class name}\\
        I&\TYPE{Interface name}\\
        T&\TYPE{Type name}\\
        m &\TYPE{Method name}\\
    \end{array}
     \;\;
     &
      \begin{array}[t]{rrl}
        \multicolumn{3}{l}{\emph{Definitions.}}\\
P &::=& \many{\mathit{IF}}\ \many{\mathit{CL}}\ \{\many{T}\ \many{x};\, s \}\\
T&::=& \TYPE{Bool} \sep \TYPE{Any} \sep I \sep \comp{\many{I}}\\
        \mathit{IF}&::=&
        \Absdisp{interface}\ I\ \Absdisp{extends}\ \many{I}\,\{\,\many{Sg}\,\}\\
        CL &::=& \Absdisp{class}\ C(\many{T}\ \many{x})\, 
\Absdisp{implements}\ \many{I}\ \{\,\many{T}\ \many{x};\{\many{T}\
\many{x}; s\};\, \many{M}\}\\
      \textsl{Sg} &::=& T\ m\ (\option{\many{T}\ \many{x}})\\
       \textsl{M} &::=& \textsl{Sg}\ \{ \many{T}\ \many{x};\ s;\Absdisp{return}\ x;\, \}\\   
e &::=& x 
\sep x.m(\many{x}) 
\sep \Absdisp{new} \ C(\many{x}) 
\sep \Absdisp{newgroup} 
\sep \Absdisp{acquire}\ I\ [\Absdisp{in}\ x]\ \Absdisp{except}\ \many{x}\\
    s&::=& 
\Absdisp{skip} 
\sep x=e 
\sep s;s 
\sep \Absdisp{if}\ x\ \{\, s\, \}\ \Absdisp{else}\, \{\, s\, \} 
\sep  \Absdisp{while}\ x \{ s \} \\
&\sep& x~\Absdisp{joins}\ x~\Absdisp{as}\ \many{I}  
\sep x\ \Absdisp{leaves}\ x ~\Absdisp{as}\ \many{I}\ \{ s \}\ \Absdisp{else}\ \{ s \}\\
&\sep& x\ \Absdisp{subtypeOf}\ I\ x\ \{ s \}\ \Absdisp{else}\ \{ s \}
     \end{array}\end{array}$$
\caption{\label{fig:syntax} Syntax of the kernel language. The type names $T$ include
  interfaces names $I$ and $\TYPE{Bool}$. Square brackets [] denotes optional elements.}
\end{figure}

\subsection{Example}
We illustrate the dynamic organization of objects in groups by an
example of software which provides text editing support (inspired by
\cite{pucella02oopsla}). This software provides two interfaces:
\Absdisp{SpellChecker} allows the spell-checking of a piece of text and
\Absdisp{Dictionary} provides functionality to update the underlying
dictionary with new words, alternate spellings, etc. Apart from an
underlying shared catalog of words, these two interfaces need not
share state and may be implemented by different classes. Let us assume
that the overall system contains several versions of
\Absdisp{Dictionary}, some of which may have an integrated
\Absdisp{SpellChecker}. Consider a class implementing a text editor
factory, which manages groups implementing these two interfaces. The
factory has two methods: \Absdisp{makeEditor} dynamically assembles
such software into a text editor group and \Absdisp{replaceDictionary}
allows the \Absdisp{Dictionary} to be dynamically replaced in such
a group. These methods may be defined as follows:

\begin{abs}
Group$\langle$SpellChecker,Dictionary$\rangle$ makeEditor() {
  Group$\langle\emptyset\rangle$ editor; SpellChecker s; Dictionary d;
  editor = newgroup; 
  d = acquire Dictionary except emptyset; 
  d subtypeOf SpellChecker ds {
     ds joins editor as Dictionary, SpellChecker; 
  } else { 
     d joins editor as Dictionary;
     s = new SpellChecker(); 
     s joins editor as SpellChecker;
  }
  return editor;
}

void replaceDictionary(Group$\langle$SpellChecker,Dictionary$\rangle$ editor, Dictionary nd){  
  Dictionary od;
  nd joins editor as Dictionary;
  od = acquire Dictionary in editor except nd;
  od leaves editor as Dictionary {skip;} else {skip;};
  return;
}
\end{abs}

The method \Absdisp{makeEditor} acquires a top-level service
\Absdisp{d} which exports the interface \Absdisp{Dictionary} (since
there is no \Absdisp{in}-clause in the \Absdisp{acquire}-expression).
If \Absdisp{d} also supports the \Absdisp{SpellChecker} interface, we
let \Absdisp{d} join the newly created group \Absdisp{editor} as
\emph{both} \Absdisp{Dictionary} and \Absdisp{SpellChecker}. Otherwise
\Absdisp{d} joins the \Absdisp{editor} group only as
\Absdisp{Dictionary}. In this case a new \Absdisp{SpellChecker} object
is created and added to the group as \Absdisp{SpellChecker}.  Remark
that we assumed the presence of several \Absdisp{Dictionary} services
in the overall system, otherwise the initial
\Absdisp{acquire}-expression may not succeed and execution could be
blocked at this point. The kernel language could be extended by a more
robust version of \Absdisp{acquire} which uses branching (similar to
\Absdisp{subtypeOf}); in fact, inside a group $g$, robustness may be
obtained by first checking for the existence of an interface $I$ in
$g$ using \Absdisp{subtypeOf} and then binding to the object or group
implementing $I$ in $g$ using \Absdisp{acquire}.

The method \Absdisp{replaceDictionary} will replace the \Absdisp{Dictionary}
service in a text editor group. First we add the new \Absdisp{Dictionary}  service
\Absdisp{nd} to the \Absdisp{editor} group and then we fetch the old
service \Absdisp{od} in the group by means
of an \Absdisp{acquire}, where the \Absdisp{except}-clause
is used to avoid binding to the new service \Absdisp{nd}. Finally the
old service  \Absdisp{od} is removed as \Absdisp{Dictionary} in the
group by a \Absdisp{leave} statement.
The example illustrates group management by joining and leaving
mechanisms as well as service discovery.


\section{A Type and Effects System}\label{sec:typing}
The language distinguishes behavior from implementations by using an
interface as a type which describes a service.  Classes are not types
in source programs. A class can implement a number of service
interfaces, so its instances can export these services to clients. A
program variable typed by an interface can refer to an instance of any
class which implements that interface. A group typed by
$\comp{\many{I}}$ exports the services described by the set $\many{I}$
of interfaces to clients, so a program variable of type $I$ may refer
to the group if $I\in\many{I}$. We denote by $\TYPE{Any}$ the
``empty'' interface, which extends no interface and declares no method
signatures.  A service described by an interface may consist of only
some of the methods defined in a class which implements the interface,
so interfaces lead to a natural notion of hiding for classes.  In
addition to the source program types used by the programmer, class
names are used to type the self-reference $\key{this}$; i.e., a class
name is used as an interface type which exports \emph{all} the methods
defined in the class.

\begin{figure}[t]
\centering \myttsize 
\renewcommand{\arraystretch}{0.9} 
$\begin{array}{c}
\nredrule{T-Var}{\Gamma \vdash x:\Gamma(x)}
\qquad

\ntyperule{T-Call}{\Gamma \vdash x: T'\quad 
\Gamma \vdash \many{x}:\many{T}\\
\textit{match}(m,\many{T}, T') \quad
\textit{retType}(T',m)= T}
{\Gamma\vdash x.m(\many{x}): T}
\qquad

\ntyperule{T-New}{\Gamma\vdash \many{x}:\textit{ptypes}(C)\quad C\prec I}
{\Gamma\vdash \key{new} \ C(\many{x}) :I }
\\[24pt]

\nredrule{T-Group}{\Gamma\vdash \key{newgroup}:\comp{\emptyset}}
\qquad

\ntyperule{T-Acquire}{\Gamma\vdash y:\comp{S}}
{\Gamma\vdash\Absdisp{acquire}\ I\ \Absdisp{in}\ y\ \Absdisp{except}\ \many{x}:I}
\qquad

\ntyperule{T-Sub}{T\prec T'\quad\Gamma\vdash  e:T}{\Gamma\vdash e:T'}
\end{array}$
\caption{\label{fig:type2} The type system for expressions. }
\end{figure}

\paragraph{Subtyping.} The subtype relation $\prec$ is defined as the
transitive closure of the extends-relation on interfaces: if $I$
extends $J'$ and $J'\prec J$ or $J'=J$, then $I\prec J$. It is
implicitly assumed that all interfaces extends $\TYPE{Any}$, so we let
$I\prec \TYPE{Any}$ for all $I$. A group type $\comp{S}$ is a subtype
of $I$ if there is some $J \in S$ such that $J \prec I$, and $\comp{S} \prec \comp{S'}$
if for all $J \in S'$ there is some $I \in S$ such that $I
\prec J$. We extend the source language subtype relation by letting a
class be a subtype of all its implemented interfaces. The reflexive
closure of $\prec$ is denoted $\preceq$.

\paragraph{Typing contexts.}
A typing context $\Gamma$ binds variable names to types. If $\Gamma$
is a typing context, $x$ a variable, and $T$ a type, we denote by
$\dom(\Gamma)$ the set of names which are bound to types in $\Gamma$
(the domain of $\Gamma$) and by $\Gamma(x)$ the type bound to $x$ in
$\Gamma$. Define the \emph{update} $\Gamma[x\mapsto T]$ of a typing
context $\Gamma$ by $\Gamma[x\mapsto T](x)=T$ and $\Gamma[x\mapsto
T](y)=\Gamma(y)$ if $y\neq x$. By extension, if $\many{x}$ and
$\many{T}$ denote lists $x_1,\ldots,x_n$ and $T_1,\ldots,T_n$, we may
write $\Gamma[\many{x}\mapsto\many{T}]$ for the typing context
$\Gamma[x_1\mapsto T_1]\ldots[x_n\mapsto T_n]$ and
$\Gamma[\many{x_1}\mapsto\many{T_1},\many{x_2}\mapsto\many{T_2}]$ for
$\Gamma[\many{x_1}\mapsto\many{T_1}][\many{x_2}\mapsto\many{T_2}]$.
For typing contexts $\Gamma_1$ and $\Gamma_2$, we define
$\Gamma_1\circ\Gamma_2$ such that
$\Gamma_1\circ\Gamma_2(x)=\Gamma_2(x)$ if $x\in\dom(\Gamma_2)$ and
$\Gamma_1\circ\Gamma_2(x)=\Gamma_1(x)$ if $x\not\in\dom(\Gamma_2)$.

For typing contexts $\Gamma_1$ and $\Gamma_2$, we define the
\emph{intersection} $\Gamma_1\cap\Gamma_2$ by
$\Gamma_1\cap\Gamma_2(x)= T$ if $T$ is the best type such
that $\Gamma_1(x)=T_1$, $\Gamma_2(x)=T_2$, and $T_1 \preceq T$ and $T_2
\preceq T$. In particular, we have $\Gamma_1\cap\Gamma_2(x)=\comp{S_1\cap S_2}$ if
$\Gamma_1(x)=\comp{S_1}$ and $\Gamma_2(x)=\comp{S_2}$.

\begin{figure}[t]
\centering \myttsize 
\renewcommand{\arraystretch}{0.9} 
$\begin{array}{c}
\nredrule{T-Skip}{\Gamma\vdash \key{skip}:\ok}
\qquad

\ntyperule{T-Assign}{\Gamma \vdash e: \Gamma(x)}{\Gamma\vdash x = e:\ok}
\qquad

\ntyperule{T-Return}
          {\Gamma \vdash s:\ok\effect{\Delta} \quad \Gamma\circ\Delta \vdash x:T}
          {\Gamma \vdash s;\Absdisp{return}\ x:T}
\qquad

\ntyperule{T-Composition}{\Gamma\vdash s:\ok\effect{\Delta_1} \quad  \Gamma\circ\Delta_1 \vdash s':\ok\effect{\Delta_2}}
{\Gamma \vdash s;s':\ok\effect{\Delta_1\circ\Delta_2}}
\\[20pt]

\ntyperule{T-Conditional}{\Gamma(x)=\TYPE{Bool}\quad
\Gamma\vdash s_1:\ok\effect{\Delta_1}  \quad \Gamma\vdash s_2:\ok\effect{\Delta_2} }
{\Gamma\vdash \key{if}\ x\,\{ s_1 \}\, \key{else}\, \{ s_2\}:\ok\effect{\Delta_1\cap\Delta_2}}
\qquad

\ntyperule{T-While}{\Gamma(x)=\TYPE{Bool}\quad
\Gamma\vdash s:\ok\effect{\Delta}}
{\Gamma\vdash \key{while}\ x\,\{ s \}:\ok\effect{\Delta}}

\\[25pt]

\ntyperule{T-Join}{\nt{local}(y)\quad \Gamma(y)=\comp{S}
\quad
\Gamma(x)\prec\many{I} }
{\Gamma\vdash x~\key{joins}\ y~\key{as}\ \many{I}
  :\ok\effect{y\mapsto\comp{S\cup \many{I}}}}
\qquad

\ntyperule{T-Leave}{\Gamma(x)\prec \many{I} \quad \Gamma(y)=\comp{S}\\
\Gamma\vdash s_1:\ok\effect{\Delta_1} 
\quad \Gamma\vdash s_2:\ok\effect{\Delta_2}}
{\Gamma \vdash x\ \Absdisp{leaves}\ y ~\Absdisp{as}\ \many{I}\ \{ s_1
\}\ \Absdisp{else}\ \{ s_2 \}:\ok\effect{\Delta_1\cap\Delta_2}}
\\[25pt]

\ntyperule{T-Inspect}{\Gamma(x)=\comp{S}\quad y\not\in\dom(\Gamma)\\
\Gamma[y\mapsto\comp{S\cup\{I\}}]\vdash s_1:\ok\effect{\Delta_1} \quad
\Gamma\vdash s_2:\ok\effect{\Delta_2} }
{\Gamma \vdash x\ \Absdisp{subtypeOf}\ I\ y\ \{ s_1\}\ \Absdisp{else}\ \{
  s_2 \}:\ok\effect{\Delta_1\cap\Delta_2}}

\qquad

\nspacetyperule{T-Method}
{\Gamma'=\Gamma[\many{x}\mapsto\many{T},\many{x'}\mapsto\many{T'}]\\
\Gamma' \vdash s;\Absdisp{return}\ e:T''\effect{\Delta}}
{\Gamma \vdash T''\ m\ (\many{T}\ \many{x})\{\many{T'}\ \many{x'}; s;
  \key{return}\ x \}:\ok}

\\[25pt]

\nspacetyperule{T-Class}
{\Gamma[\textrm{this}\mapsto C,
\many{x_2}\mapsto\many{T_2}]\vdash
\many{M}:\ok\\
C\prec \many{I}\quad \Gamma[\textrm{this}\mapsto C, \many{x_2}\mapsto\many{T_2}. \many{x_1}\mapsto\many{T_1},\many{x_3}\mapsto\many{T_3}]\vdash s:\ok\effect{\Delta}}
{\Gamma \vdash \key{class}\ C(\many{T_1}\ \many{x_1})\ \key{implements}\
  \many{I}\ \{\many{T_2}\ \many{x_2}; \{\many{T_3}\ \many{x_3}; s\}; \many{M}\}:\ok}
\qquad

\nspacetyperule{T-Program}{\Gamma[\many{x}\mapsto\many{T}] \vdash s:\ok\effect{\Delta}\\
\forall CL\in\many{CL}\cdot \Gamma \vdash CL:\ok}
{\Gamma \vdash 
\many{IF}\ \many{CL}\ \{\many{T}\ \many{x}; s\}:\ok}
\end{array}$
\caption{\label{fig:type3} The type and effect system for statements, methods,
  classes, and programs.}
\end{figure}

\paragraph{The Type and Effect System.}
Programs in the kernel language are analyzed using a type and effect
system (e.g., \cite{talpin92effect,amtoft99book,lucassen88popl}). The
inference rules for expressions are given in Figure~\ref{fig:type2}
and for statements, methods, classes, and programs in
Figure~\ref{fig:type3}.

\emph{Expressions} are typed by the rules in Figure~\ref{fig:type2}.
Let $\Gamma$ be a typing context.  A typing judgment $\Gamma \vdash
e:T$ states that the expression $e$ has the type $T$ if the variables
in $e$ are typed according to $\Gamma$. By \textsc{T-Var}, variables
must be typed in $\Gamma$. Method calls to a method $m$ on a variable
$x$ are typed to $T$ if $x$ has the (interface) type $T'$ such that
the types $\many{T}$ of the actual parameters $\many{x}$ give a match
for $m$ in $T'$ with parameter types $\many{T}$ and the declared
return type of $m$ in $T'$ is $T$. In \textsc{T-New}, \key{new} $C$
has type $I$ if the types of the actual parameters to the class
constructor can be typed to the declared types of the formal
parameters of the class, by means of the auxiliary function
$\mathit{ptypes}$, and the class implements $I$, expressed by $C\prec
I$.  We omit the definitions of the auxiliary functions
$\mathit{match}$ and $\mathit{retType}$ here, these are
straightforward lookup functions on the program's interface table
which perform the matching and retrieve the return type of a method in
a class, respectively.  Similarly, $\mathit{ptypes}$ retrieves the
types of the formal parameters to a class in the program's class
table.
By \textsc{T-Group}, a new group has the empty group type
(with no exported interfaces). By \textsc{T-Acquire}, service
discovery has the obvious type, if successful. The premise of the rule
is omitted if the statement has no \Absdisp{in}-clause. Rule
\textsc{T-Sub} captures 
subtyping in the type system.

\emph{Statements} are typed by the rules in Figure~\ref{fig:type3}.
Let $\Gamma$ and $\Delta$ be typing contexts.  A typing judgment
$\Gamma \vdash s:\ok\effect{\Delta}$ expresses that the statement $s$
is well-typed if the variables in $s$ are typed according to $\Gamma$
and that the typing context for further analysis should be modified
according to the \emph{effect} $\Delta$. Empty effects are omitted in
the presentation of the rules.  The typing of statements
$\Absdisp{skip}$ and $x=e$ are standard. These judgments have no
effects. The statement $\Absdisp{return}\ x$ has a return type and is
typed in the effect of typing the statements of the method body. The
use of effects can be seen in rule \textsc{T-Composition}, where the
second statement is type checked in the typing context modified by the
effect of analyzing the first statement, and the effects are
accumulated in the conclusion of the rule. Rules
\textsc{T-Conditional} and \textsc{T-While} propagate effects from the
subexpressions; in the case of \textsc{T-Conditional} the resulting
effect is approximated by taking the intersection of the effects of
the branches. By \textsc{T-Join}, when an object joins a group $y$ and
contributes interfaces $\many{I}$ to $y$, the effect is that the type
of $y$ is extended with the interfaces $\many{I}$.  Note the
requirement $\mathit{local}(y)$, which expresses that $y$ must be a
local variable in the scope of the method being analyzed.  (We omit
the definition, which is again a lookup in the class table of the
program). Without this restriction, a field could dynamically extend
its type, resulting in an unsound system; e.g., an assignment
\Absdisp{f=e} in a statically well-typed method could become unsound
if the type of \Absdisp{f} were extended. However extending the type
\Absdisp{T} of a local variable which copies the value of \Absdisp{f}
to a type \Absdisp{T'} and assigning the result back to a field
\Absdisp{f'} is allowed, as \Absdisp{f'} would need to be of the
extended type \Absdisp{T'} and \Absdisp{f} would remain of type
\Absdisp{T} as required by the other method. (For comparison, the
needed restriction to local variables is handled differently in the
query statement \Absdisp{subtypeOf}, which introduces a fresh local
variable.) Rule \textsc{T-Leave} shows that leaving a group has no
effect on the typing context, and the effects of the two branches are
treated as for the conditional. Rule \textsc{T-Inspect} shows how the
typing context is extended with a new variable $y$ which extends the
type of the group $x$ for the scope of the branch $s_1$. The overall
effect is again the intersection of the effects of the two branches.

Programs, classes, and methods are typed in the standard way.  Methods
do not have effects, which reflects that effects are constrained to
local variables inside methods.  Likewise, classes and programs do not
have effects. (For simplicity, the standard type checking of interface
declarations is omitted in the presentation.) The body of a class
constructor and the main method of a program may have the same effects
as the body of a method.


\section{Operational Semantics}
\label{sec:opsem}
\begin{figure}[t!]
\myttsize
$$\begin{array}{cl}
      \begin{array}[t]{c@{\,:\,}l}
         \multicolumn{2}{l}{\emph{Syntactic Categories.}}\\
       g & \TYPE{Group name}\\
        o & \TYPE{Object name}
     \end{array}
      \;\;
      &
      \begin{array}[t]{rrl}
        \multicolumn{3}{l}{\emph{Definitions.}}\\
\nt{cn} &::=& \epsilon \sep \nt{grp} \sep \nt{obj} \sep \nt{cn}~\nt{cn}\\
\nt{grp} &::=& g(\nt{export})\\
\nt{export}&::=& \{o:I\} \sep \nt{export}\cup\nt{export}\\ 
        \nt{obj} &::=& o(\sigma, \rho)\\
\rho & ::=& \nt{idle} \sep \nt{proc} \sep \nt{proc};\rho\\
        \nt{proc} &::=& m\{\sigma | sr\} \sep \textit{error}\\
        \sigma & ::= & x\mapsto \langle T, v\rangle \sep \sigma\circ\sigma\\
     e &::=& 
      \Absdisp{wait}(o,m) \sep \ldots\\ 
      sr&::=&s \sep s; \Absdisp{return}\ x;\\
      v&::=& o \sep g \sep \mathrm{true} \sep \mathrm{false}
     \end{array}\end{array}$$
\caption{\label{fig:runtimesyntax}The runtime syntax,
extending the language syntax for expressions $e$ and statements $s$.}
\end{figure}

The runtime syntax is given in Figure~\ref{fig:runtimesyntax}.  A
runtime configuration $cn$ is either the empty configuration
$\varepsilon$ or it consists of objects $\nt{obj}$ and groups
$\nt{grp}$. Groups $\nt{grp}$ have an identity $g$ and contain a set
$\nt{export}$ of interfaces $I$ associated with the objects $o$
implementing them.  Objects $\nt{obj}$ have an identity $o$, a state
$\sigma$, and a stack $\rho$ of processes $\nt{proc}$. When an object
has processes to execute, it executes the process at the top of its
stack.  The stack grows with self-calls and shrinks at method
returns. The empty stack is denoted $\nt{idle}$.  A state $\sigma$
maps program variables $x$ to their types $T$ and values $v$.  A
process $\nt{proc}$ can be $\nt{error}$ or it has a local state
$\sigma$ and a sequence $s; \key{return}\ x;$ of statements to be
executed. The expression $\key{wait}(o,m)$ encodes a \emph{lock},
expressing that the object is waiting for the return value of method
$m$ in another object $o$ (or on an auxiliary self-call).  Values $v$
include object and group names, and Booleans.

The operational semantics is given by rules in the style of SOS
\cite{plotkin04jlap}, reflecting small-step semantics.  Each rule
describes one step in the execution of an object.  Concurrent
execution is given by standard SOS context and concurrency rules (not
shown here), and we assume associative and commutative matching over
configurations (as in rewriting logic \cite{maude-book}).  Thus
objects execute concurrently, with the following exceptions: The rule
for synchronous remote call ({\sc{Call1}}) refers to both the caller
and callee objects and therefore the two objects must
\emph{synchronize} and the caller will be blocked by the
\Absdisp{wait} statement.  Furthermore rules involving an object and a
group will lock the group in question, thereby disallowing concurrent
execution of other objects involving the same group.  This is crucial
in the \textsc{Join} and \textsc{Leave1} rules for \Absdisp{joins} and
\Absdisp{leaves}, which may actually modify the group.

We define the lookup of a program variable $x$ in a state $\sigma$ by
$\eval{x}{\sigma}=\langle T, v\rangle$, with the projections
$\evalT{x}{\sigma}= T$ and $\evalV{x}{\sigma}=v$. Thus, for a state
$\sigma$, $\sigma^T$ gives the associated mapping of program variables
to their types and $\sigma^V$ the mapping of program variables to
their values.
The {\sc{Skip}} rule is standard
and states that a skip has no effect.
The effect of assignment is divided into two rules, {\sc{Assign1}} for
local variables, updating $l$, and {\sc{Assign2}} for fields, updating
$a$.
In the rule {\sc{New-Group}}, a globally unique group identifier is
found by $\nt{fresh}(g)$. Then an empty group with this identifier is
added to the configuration.
The two rules {\sc{Cond1}}  and {\sc{Cond2}} handle the two cases of
the conditional statement.

\emph{Method calls} are handled by {\sc{Call1}} for calls to other
objects, {\sc{Call2}} for self calls, and {\sc{Call3}} for calls to
groups. When a call is made to another object in {\sc{Call1}}, the
called object must be in an \textit{idle} state. The caller blocks
until the generated \Absdisp{wait} statement can be executed. In the
\Absdisp{wait} statement, the callee and method name are recorded,
which allows the runtime type system to infer the proper type of the
return value from method $m$ in the proper class.  Let
$\nt{bind}(m,C,\many{v})$ denote the process resulting from the
activation of method $m$ in $C$, in which $l$ maps the parameters of
$m$ to their declared types and values $\many{v}$, and the local
variables to their declared types and default values.  The callee gets
the process $\textit{bind}(m,C,\evalV{\many{y}}{(a\circ l)})$, where
$C$ is the class of the callee, pushed onto its process stack
$\rho$. With self calls in {\sc{Call2}}, the process stack cannot be
idle, but a \Absdisp{wait} statement replaces the call statement and
an instance of the called method is pushed to the stack. In
{\sc{Call3}}, a call to a group is reduced to a call to a group or an
object \emph{inside} the callee which exports an appropriate interface
to the group. By appropriate we mean that the called method is
supported by the interface (formally, $m\in\nt{mtd}(I)$).
{\sc{Return1}} handles returns from remote calls. Here the blocking
\Absdisp{wait} statement is replaced by the returned value. Returns
from self calls are handled in a similar way by the {\sc{Return2}}
rule. (Remark that the generalization to concurrent objects with
asynchronous calls and futures is straightforward as in
\cite{johnsen10fmco,DBLP:conf/birthday/ClarkeJO10} whereas the
extension to multi-threaded programs would require re-entrant lock as
in \cite{igarashi01}).

The \Absdisp{new} statement is handled by the {\sc{New-Object}} rule,
where $\nt{fresh}(o',C)$ asserts that $o'$ is a new name in the global
configuration such that $\nt{classOf}(o')=C$.  An object with this
name is created. The mapping $\nt{atts}(C,\many{v})$ maps the declared
fields of class $C$ to their declared types and default values,
$\nt{this}$ to $C$, and the class parameters to declared types and
actual values. The process $\nt{init}(C)$ corresponds to the
init-block of $C$, which instantiates local variables to their
declared types and default values.  The process of the new object is
the initial process of its class.  Note that an init-block is executed
independently from the creator, so it may trigger \emph{active
  behavior}; for instance, the init-block can call a run method.

The rule {\sc{Join}} extends the knowledge of a group with the new
interfaces from the object's perspective and correspondingly extends
the $\nt{exports}$ set from the group's perspective.  Service
discovery is handled by the {\sc{Acquire}} rule. The \Absdisp{acquire}
expression is replaced by a value $v$, which is an object or group
identifier satisfying the \Absdisp{in} and \Absdisp{except}
clauses. If the \Absdisp{in} clause is omitted from the expression,
then the premise $ \evalV{y}{(a\circ l)}=g $ is omitted from the
rule. Note that this rule will block if no matching object or group
exists. This could be solved by either returning \Absdisp{null} (by
means of a global check) or by adding an \Absdisp{else} branch similar
to those in {\sc{Query1}} and {\sc{Query2}}.  Within the kernel
language, the existence of a matching object or group inside a group
can be checked using the query mechanisms.

The \Absdisp{leaves} statement is handled by the rules {\sc{Leave1}}
for a successful leave and {\sc{Leave2}} for an unsuccessful one.  A
group or object $x$ may leave a group successfully if the group
provides the same interface support without $x$.  To determine this,
we use the function $\nt{intf}(\nt{export})$ which returns a set
containing the interfaces of all the pairs in $export$, removing
redundant information.  An entry is redundant if a subtype of the
entry is present in the set.  The type of the group does not change by
a \Absdisp{leaves} statement and hence the object does not need to
update information about the group. The branches $s_1$ or $s_2$ are
chosen depending on the success.  The rules {\sc{Query1}} and
{\sc{Query2}} handle the branching statement that checks if a group
exports a given interface. If the test succeeds then a fresh variable
$y$ is introduced and is only visible in $s_1$. The type of this
variable is the union of what the current object already knew about
the group and the new information $I$. If the test fails the $s_2$
branch is chosen by {\sc{Query2}}.

\emph{The initial state.}
For a program $P= \many{IF}\ \many{CL}\ \{\many{T}\ \many{x}; s\}$, we
define the initial state to be $ o(\epsilon, \textit{main}\{\many{x}
\mapsto \langle \many{T},\textit{default}(\many{T})\rangle|s; \})$
where $o$ is such that $\textit{fresh}(o,Main)$.

\begin{figure*}[p]
\myttsize
$$\begin{array}{c}

\nredrule{Skip}
{o(a, m\{l \mid \key{skip};sr\};\rho)\\\to o(a, m\{l \mid sr\};\rho)}

\quad

\ntyperule{Assign1}{x\in\,\textit{dom}(l)\\\evalT{x}{l}=T\quad
  \evalV{y}{(a\circ l)} =v}
{o(a, m\{l \mid x=y;sr\};\rho) \to\\
o(a, m\{l[x\mapsto\langle T,v\rangle] \mid sr \};\rho)}
\quad
\ntyperule{Assign2}{x\notin\,\textit{dom}(l) \\\evalT{x}{a}=T\quad
  \evalV{y}{(a\circ l)} =v}
{o(a,  m\{l \mid x=y;sr\};\rho) \to\\
o(a[x\mapsto\langle T,v\rangle],  m\{l \mid sr \};\rho)}

\quad

\ntyperule{New-Group}{\textit{fresh}(g)}{o(a, m\{l \mid x=\key{newgroup};sr\};\rho)\\
\to o(a, m\{l \mid x=g ;sr\};\rho)\ g(\emptyset)}

\\\\

\ntyperule{Cond1}{\evalV{x}{(a\circ l)}}
{o(a, m\{l|\key{if}\ x\  \{s_1\}\ \key{else}\ \{s_2\};sr\};\rho)\\
\to (o(a, m\{l|s_1;sr\};\rho)}
\quad
\ntyperule{Cond2}{\neg \evalV{x}{(a\circ l)}}
{o(a, m\{l|\key{if}\ x\  \{s_1\}\ \key{else}\ \{s_2\};sr\};\rho)\\
\to o(a, m\{l|s_2;sr\};\rho)}
\quad

\nredrule{While}
{o(a, m\{l \mid \key{while}\ x\ \{s_1\};sr\};\rho)\\
\to o(a, m\{l \mid \key{if}\ x\  \{s_1\ ;\ \key{while}\ x\ \{s_1\}\}\\
\qquad\qquad\key{else}\ \{\key{skip}\}\ ; sr\};\rho)}

\\\\
\ntyperule{Call1}
{\evalV{y}{(a\circ l)} = o' \quad \textit{classOf}(o')=C\\
\textit{pr}=\textit{bind}(m,C,\evalV{\many{y}}{(a\circ l)})}
{o(a, m\{l\mid x=y.m(\many{y});sr\};\rho)\
  o'(a', \textit{idle}) \to\\  
  o(a, m\{l \mid x=\key{wait}(o',m);sr\};\rho)\
  o'(a', pr)}
\;\;\;\;

\ntyperule{Call2}
{\evalV{y}{(a\circ l)} = o \quad \textit{classOf}(o)=C\\
\textit{pr}=\textit{bind}(m,C,\evalV{\many{y}}{(a\circ l)})}
{o(a, m\{l\mid x=y.m(\many{y});sr\};\rho) \to\\  
  o(a, pr;m\{l \mid x=\key{wait}(o,m);sr\};\rho)}

\;\;\;\;

\ntyperule{Call3}
{\evalV{y}{(a\circ l)}=g\\
v:I\in\nt{exports}\quad
m \in \textit{mtd}(I) }
{o(a, m\{l \mid x=y.m(\many{y});sr\};\rho)\ g(\nt{exports})\\
\to
o(a, m\{l \mid x=v.m(\many{y});sr\};\rho) \ g(\nt{exports})}
\\\\

\ntyperule{Return1}
{\evalV{x}{(a\circ l)} = v \quad \rho = \nt{idle}}
{o(a, m\{l\mid \key{return}\ x;\};\rho)\\
o'(a',  m'\{l' \mid y=\key{wait}(o,m);sr\};\rho')\\
\to o(a, \rho)\
 o'(a',  m'\{l'\mid y=v;sr\};\rho')}

\quad

\ntyperule{Return2}
{\evalV{x}{(a\circ l)} = v}
{o(a, m\{l\mid \key{return}\ x;\};\hspace{30pt}\\
\hspace{30pt}m'\{l' \mid y=\key{wait}(o,m);sr\};\rho)\\
\to o(a, m'\{l'\mid y=v;sr\};\rho)}

\quad

\ntyperule{New-Object}{
\textit{fresh}(o', C) \quad
\textit{pr}=\textit{init}(C)\\
  a'=\textit{atts}(C,\evalV{\many{x}}{(a\circ l)})}
{o(a, m\{l| x=\key{new}\ C(\many{x});sr\};\rho) 
\\\to o(a, m\{l|x=o';sr\};\rho)\
  o'(a', \textit{pr}) }
\\\\

\ntyperule{Join}{
\evalV{x}{(a\circ l)} = v \quad
\eval{y}{l}=\langle \comp{S},g\rangle\\
T = \comp{S\cup \many{I}} \quad
\nt{exports}'=\bigcup_{I\in \many{I}}\{v:I\}\cup\nt{exports}}
{o(a, m\{l| x\ \key{joins}\ y\
  \key{as}\ \many{I};sr\};\rho)\ g(\nt{exports}) 
\\\to o(a, m\{l[y\mapsto\langle T,g\rangle]|sr\};\rho)\ g(\nt{exports}') }
\quad

\ntyperule{Acquire}{\evalV{y}{(a\circ l)}=g \quad (v:J)\in\nt{exports}\quad J\prec I \quad
v \notin \evalV{\bar{x}}{(a\circ l)} }
{o(a, m\{l \mid x=\key{acquire}\ I\ \key{in}\ y\ \key{except}\ \bar{x};sr\};\rho)\
g(\nt{exports})\\
\to 
o(a, m\{l \mid x= v ;sr\};\rho)\ g(\nt{exports}) }
\\\\

\ntyperule{Leave1}{\evalV{y}{(a\circ l)}=g \quad
\evalV{x}{(a\circ l)} = v \\
\nt{exports}'= \nt{exports} \setminus \bigcup_{I\in \many{I}}\{v:I\} \quad
\nt{intf}(\nt{exports}) = \nt{intf}(\nt{exports}') 
}
{o(a, m\{l| x\ \key{leaves}\ y ~\key{as}\ \many{I}\ \{ s_1
\}\ \key{else}\ \{ s_2 \};sr\};\rho)\ g(\nt{exports})\\
\to o(a, m\{l|s_1;sr\};\rho) \ g(\nt{exports'})}
\quad

\ntyperule{Leave2}{\evalV{y}{(a\circ l)}=g \quad
\evalV{x}{(a\circ l)} = v \\
\nt{exports}'= \nt{exports} \setminus \bigcup_{I\in \many{I}}\{v:I\} \quad
\nt{intf}(\nt{exports}) \neq \nt{intf}(\nt{exports}') 
}
{o(a, m\{l| x\ \key{leaves}\ y ~\key{as}\ \many{I}\ \{ s_1
\}\ \key{else}\ \{ s_2 \};sr\};\rho)\ g(\nt{exports})\\
\to o(a, m\{l|s_2;sr\};\rho) \ g(\nt{exports})}

\\\\
\ntyperule{Query1}{
y\not\in\dom(a\circ l)\quad \eval{x}{a \circ l}=\langle \comp{S},g\rangle\quad 
  o':J\in\nt{exports}\quad J\prec I}
{o(a, m\{l| x\ \key{subtypeOf}\ I\ y\ \{s_1\}\ \key{else}\ \{s_2\};sr\};\rho)\\ g(\nt{exports})\\
\to o(a, m\{l[y\mapsto\langle \comp{S\cup\{I\},g\rangle}]|s_1;sr\};\rho) \ g(\nt{exports})}
\quad

\ntyperule{Query2}{\evalV{x}{(a\circ l)}=g\quad
\comp{\nt{intf}(\nt{exports})}\not\prec I}
{o(a, m\{l| x\ \key{subtypeOf}\ I\ y\ \{s_1\}\ \key{else}\ \{s_2\};sr\};\rho) \\ g(\nt{exports})\\
\to o(a, m\{l|s_2;sr\};\rho) \ g(\nt{exports})}

\end{array}$$
\caption{\label{fig:sem1}The operational semantics.}
\end{figure*}


\section{Type Safety}
\label{sec:typesafe}
This section extends the type system of Section~\ref{sec:typing} to runtime configurations
and shows that the execution of well-typed programs remains well-typed.

\subsection{Well-Typed Configurations}
The extension of the type system to runtime configurations is given in
Figure~\ref{fig:typeRuntime}.  The typing context $\Gamma$ stores the
types of all constant values (object and group identities) at
runtime. By \textsc{RTT-Config}, a configuration is well-typed if all
objects and groups are well-typed. By \textsc{RTT-Group}, a group is
well-typed if all the objects which export interfaces through the
group implement these interfaces (checked by \textsc{RTT-Exps} and
\textsc{RTT-Exp}). By \textsc{RTT-Object}, an object is well-typed if
its class is its type in $\Gamma$ and its state and stack are
well-typed in the context of the types of the fields. Substitutions
(the state of fields and local variables) are checked by
\textsc{RTT-Subs} and \textsc{RTT-Sub}. The stack is well-typed by
\textsc{RTT-Stack} if all its processes are well-typed by
\textsc{RTT-Proc}; i.e., the state of local variables and the method
body $sr$ are well-typed.  Observe that due to the query-mechanism of
the language, the types of program variables in two processes which
stem from activations of the same method, may differ at runtime. For
this reason, the typing context used for typing runtime configurations
cannot rely on the statically declared types of program
variables. This explains why \textsc{RTT-Proc} extends $\Gamma$ with
the \emph{locally stored typing information} $l^T$ to type check $l^V$
and $sr$. The effects of the static type system are not needed here,
as they are reflected by how the operational semantics updates this
local type information. For consistency in the presentation, the
typing of fields is represented in the same way, although these types
are not altered by the execution.  The rules from the static type
checking are reused as appropriate.

\begin{figure}[t]
\centering \myttsize 
\renewcommand{\arraystretch}{0.9} 
$\begin{array}{c}
\nredrule{RTT-Empty}
         {\Gamma \vdash \epsilon:\ok}
\quad

\nredrule{RTT-Idle}
         {\Gamma \vdash \key{idle}:\ok}

\quad

\nredrule{RTT-Wait}
{\Gamma \vdash \Absdisp{wait}(o,m):\textit{retType}(\textit{classOf}(o),m)}

\quad

\nredrule{RTT-Def}{\Gamma \vdash \textit{default}(T):T}
\\\\

\ntyperule{RTT-Config}
          {\Gamma \vdash cn:\ok \quad \Gamma \vdash cn':\ok}
          {\Gamma\vdash cn\ cn':\ok}

\;\quad

\ntyperule{RTT-Group}
{\Gamma\vdash \nt{exports}:\Gamma(g)}
{\Gamma\vdash g(\nt{exports}):\ok}

\;\quad

\ntyperule{RTT-Exp}
{I\in S \quad \Gamma(o)\prec I}
{\Gamma\vdash o:I:\comp{S} }

\;\quad

\ntyperule{RTT-Sub}
          {\Gamma \vdash v: \Gamma(x)}
          {\Gamma \vdash x \mapsto v:\ok}

\;\quad

\ntyperule{RTT-Subs}
          {\Gamma \vdash a:\ok \quad \Gamma \vdash a':\ok }
          {\Gamma \vdash a \circ a':\ok }

\\\\

\ntyperule{RTT-Object}
{\Gamma' = \Gamma \circ a^T\quad
\Gamma' \vdash a^V:\ok \\
\textit{classOf}(o) = \Gamma(o) \quad
\Gamma' \vdash \rho:\ok}
        {\Gamma \vdash o(a,\rho):\ok }

\hfill

\ntyperule{RTT-Exps}
{\Gamma\vdash \nt{exports}:\comp{S} \\ \Gamma\vdash \nt{exports}':\comp{S}}
{\Gamma\vdash \nt{exports}\cup \nt{exports}':\comp{S}}

\hfill

\ntyperule{RTT-Proc}
{\Gamma' = \Gamma\circ l^T
\quad 
\Gamma(this) = C 
\\
\Gamma' \vdash l^V:\ok  
\quad  
\Gamma' \vdash sr:\textit{retType}(C,m)
}
{\Gamma \vdash m \{ l | sr;\}:\ok}

\hfill

\ntyperule{RTT-Stack}
{\Gamma\vdash \nt{proc}:\ok\\\Gamma\vdash \rho:\ok}
{\Gamma\vdash \nt{proc};\rho:\ok}

\end{array}$ 
\caption{\label{fig:typeRuntime} The runtime type system.}
\end{figure}

\subsection{Subject Reduction}
\newtheorem{lemma}{Lemma}
\newtheorem{theorem}{Theorem}
\newenvironment{proof}{\par\noindent\emph{Proof.}}{\hfill$\square$}

The type system guarantees that the type of \emph{fields}
in an object never changes at runtime (in particular, recall the
  restriction $\textit{local}(y)$ in rule \textsc{T-Join}). This
allows us to establish in Lemma~\ref{lemma1} from the static typing of
methods in well-typed programs that method binding, if successful,
results in a well-typed process at runtime. To show that the
\key{error} process cannot occur in the execution of well-typed
programs, it suffices to show that substitutions are always well-typed.
Lemma~\ref{lemma2} shows that this is the case for the initial
configuration and Lemma~\ref{lemma3} shows that one execution step
preserves runtime well-typedness. Together, these lemmas establish a
subject reduction theorem for the language, expressing that
well-typedness is preserved during the execution of well-typed
programs and in particular that method binding always succeeds. Here,
$\stackrel{*}{\to}$ denotes the reflexive and transitive closure of
the reduction relation $\to$.

\begin{lemma}\label{lemma1}
Assume that a well-typed program has a class $C$ which defines a method
$m$ with formal para\-meters $\many{x}$ of type $\many{T}$ and return
type $T$.  Let $o$ be
an object such that $\textit{classOf}(o)=C$ and $\Gamma\vdash
o(a,\rho):\ok$. If $\Gamma\vdash \many{v}:\many{T}$, then
$\Gamma\circ a^T\vdash
\textit{bind}(m,C,\many{v}):T$.
\end{lemma}

\begin{lemma}\label{lemma2}
  Let $P$ be a program such that $\Gamma\vdash P:\ok$ and let
  $\nt{cn}$ be the initial state of $P$.
Then $\Gamma\vdash \nt{cn}:\ok$.
\end{lemma}

\begin{lemma}\label{lemma3}
If $\Gamma\vdash cn: \ok$ and $cn\to cn'$ then there is a $\Gamma'$ such
that $\Gamma'\vdash cn':\ok$ and $\Gamma\subseteq\Gamma'$.
\end{lemma}

\begin{theorem}[Subject reduction]
Let $\Gamma\vdash P$ and let $cn$ be the initial runtime state of
$P$. If $cn \stackrel{*}{\to} cn'$ then there is a $\Gamma'$ such
that $\Gamma'\vdash cn':\ok$ and $\Gamma\subseteq\Gamma'$.
\end{theorem}


\section{Related Work}

\label{sec:related}
Object orientation is well-suited for designing small units which
encapsulate state with behavior, but does not directly address the
organization of more complex software units with rich interfaces. 
Two approaches to building flexible and adaptive complex software
systems involve, independently, object groups and service discovery.
Our work unifies these two approaches in a formal, type-safe setting.


The most common use of object groups is to provide replicated services in order to
offer better fault tolerance. Communication to elements of a group is via multicast.
This idea originated in the Amoeba operating system~\cite{DBLP:journals/dse/KaashoekTV93}.
The component model Jgroup/ARM~\cite{DBLP:journals/spe/MelingMHB08} adopts this idea to provide
autonomous replication management using  distributed object groups. 
In this setting, members of a group  maintain a replicated state for reasons of consistency.
The ProActive active object programming model~\cite{DBLP:conf/java/BaduelBC02} supports abstractions for object groups,
which enable group communication---via method call---and various means for synchronizing on the results of such method calls,
such as wait-for-one and wait-for-all. ProActive is formalized in 
Caromel and Henrio's Theory of Distributed Objects~\cite{DBLP:books/daglib/0012826}. 
These notions of group differ from ours in two respects. Firstly, in these approaches communication with groups is via multicast, 
whereas in our approach each message will be delivered to exactly one object, and
secondly, in the formal theory, groups are fixed upon creation. Furthermore, there is no notion of service 
discovery associated with groups.

Object groups have been investigated as a modularization unit for
objects which is complementary to components. Groups meet the needs of
organizing and describing the statics and dynamics of networks of
collaborating objects~\cite{lea93groups}; groups can have many threads
of control, they support roles (or interfaces), and objects may
dynamically join and leave groups. Lea~\cite{lea93groups} presents a
number of common usages for groups and discusses their design
possibilities, inspired from CORBA. 
Groups have been used to provide an abstraction akin to a notion of component.
For example, in Oracle Siebel 8.2~\cite{siebel},
 groups are used as units of deployment, units of monitoring, and units of control
when deploying and operating  components on Siebel servers.
Our approach abstracts from most of these details, though groups are treated as
first class entities in our calculus.

Another early work on groups is
ActorSpaces~\cite{agha93ppopp}, which combine Actors with Linda's
pattern matching facility, allowing both one-to-one communication,
multicast, and querying. Unlike our approach, groups in ActorSpaces
are intensional: all actors with the same interface belong to the same
group. Furthermore ActorSpaces support broadcast communication to a
group, which has not been considered in this paper as it would
differentiate communication with an object and with a group. Compared to
our paper, these works do not give a formalization of group behavior
or discuss typing.

Object groups have further been used for coordination purposes. For
example, CoLaS \cite{cruz99coord} is a coordination model based on
groups in which objects may join and leave groups. CoLaS goes beyond
the model in our paper by allowing very intrusive coordination of
message delivery based on a coordinator state.  In our model, the
groups don't have any state beyond the state of their objects. Similar
to our model, objects enroll to group roles (similar to
interfaces). However, unlike our model objects may leave a group at
any time, and the coordinator may access the state of
participants. The model is implemented in Smalltalk and neither
formalization nor typing is discussed \cite{cruz99coord}.
Concurrent object groups have also been proposed to define
collaborating objects with a single thread of control in programming
and modeling languages \cite{schaefer10ecoop,johnsen10fmco}.
Concurrent object groups do not have identity and function as runtime
restrictions on concurrency rather than as a linguistic concept.

Microsoft's Component Object Model (COM) supports querying a component
to check whether it supports a specific interface, similar to the
query-mechanism considered in this paper. A component in COM may also
have several interfaces, which are independent of each other. In
contrast to the model presented in our paper, COM is not
object-oriented and the interfaces of a component are stable (i.e.,
they do not change).  COM has proven difficult to formalize; Pucella
develops $\lambda^{COM}$ \cite{pucella02oopsla}, a typed
$\lambda$-calculus which addresses COM components in terms of their
interfaces, and discusses extensions to the calculus to capture
subtyping, querying for interfaces, and aggregation.

 A wide range of service discovery
mechanisms exist~\cite{DBLP:conf/icsoc/Hasselmeyer05}.  
The programming language
AmbientTalk~\cite{DBLP:conf/ecoop/DedeckerCMDM06} has built-in service
discovery mechanisms, integrated in an object-oriented language with
asynchronous method calls and futures. In contrast to our work, AmbientTalk is an untyped
language, and lacks any compile time guarantees.
Various works formalise the notion of
service discovery~\cite{DBLP:journals/entcs/LapadulaPT08}, but they
often do so in a formalism quite far removed from the standard setting
in which a program using service discovery would be written, namely,
an object-oriented setting.  For example, Fiadeiro et
al.'s\cite{DBLP:journals/fac/FiadeiroLB11} model of service discovery
and binding takes an algebraic and graph-theoretic approach, but it
lacks the concise operational notion of service discovery formalized
in our model.  No type system is presented either.

Some systems work has been done that combines groups and service discovery mechanisms,
such as group-based service discovery mechanisms
in mobile ad-hoc networks~\cite{DBLP:conf/mwcn/ChakrabortyJYF02,DBLP:journals/cn/GaoWYY06}.
In a sense our approach provides language-based abstractions for a mechanism like this,
except that ours also is  tied to interface types to ensure type soundness
and includes a notion of exclusion to filter matched services. 

Our earlier work~\cite{DBLP:conf/birthday/ClarkeJO10} enabled
objects to advertise and retract interfaces to which other objects
could bind, using a primitive service discovery mechanism. A group
mechanism was also investigated as a way of providing structure to the
services. In that work services were equated with single objects,
whereas in the present work a group service is a collection of
objects exporting their interfaces. In particular, this means that the
type of a group can change over time as it comes to support more
functionality.

The key differences with most of the discussed works is that the model
in this paper remains within the object-oriented approach,
multiple groups may implement an advertised service in different ways,
and our formalism offers a transparent group-based service discovery
mechanism with primitive exclusion policies.  Furthermore, our notion
of groups has an implicit and dynamically changing interface.

\section{Conclusion}
\label{sec:conc}

The paper has proposed a formal model for adaptive service-oriented
systems, based on a notion of object-oriented groups.  We develop a
kernel object-oriented language in which groups are first-class
citizens in the sense that they may play the role of objects; i.e., a
reference typed by an interface may refer to an object or to a group.
A main advantage is that one may collect several objects into a group,
thereby obtaining a rich interface reflecting a complex service, which
can be seen as a single object from the outside.  Although objects in
our language are restricted to executing one method activation at the
time, a group may serve many clients at the same time due to inner
concurrency.

In contrast to objects, groups may dynamically add support for an
increasing number of interfaces.  The formation of groups is dynamic;
\emph{join} and \emph{leave} primitives in the kernel language allow
the migration of services provided by objects and inner groups as well
as software upgrade, provided that interfaces are not removed from a
group.  An object or group may be part of several groups at the same
time.  This gives a very flexible notion of group.

Adaptive object groups are combined with service discovery by means of
\emph{acquire} and \emph{subtypeOf} constructs in the kernel language,
which allow a programmer to discover services in an open and unknown
environment or in a known group, and to query interface support of a
given object or group.  These mechanisms are formalized in a general
object-oriented setting, based on experiences from a prototype Maude
\cite{maude-book} implementation of the group and service discovery
primitives.  The presented model provides expressive mechanisms for
adaptive services in the setting of object-oriented programming with
modest conceptual additions.  We have developed an operational
semantics and type and effects system for the kernel language, and
show the soundness of the approach by a proof of type-safety.

The combination of features proposed in this paper suggests that our
notion of a group can be made into a powerful programming concept.
The work presented in this paper may be further extended in a number
of directions. The overall goal of our work is to study an integration
of service-oriented and object-oriented paradigms based on a formal
foundation.  In future work, we plan to extend the proposed kernel
language to multi-thread concurrency and study in more detail how
different usages of object groups such as replication, resource, and
access groups (see, e.g., \cite{lea93groups}) may be captured using
the proposed primitives. It is also interesting to study the
integration into the kernel language of more service-oriented concepts
such as for example error propagation and handling, as well as
high-level group management operations such as group aggregation.


\bibliographystyle{eptcs}
\bibliography{refs}


\end{document}